\documentclass[12pt]{article}
\usepackage{fullpage,amssymb,amsmath}
\usepackage[dvips]{epsfig}

\def\##1{{\bf #1}}
\def\=#1{\underline{\underline{#1}}}

\def\+#1{\underline{\bf #1}}
\def\*#1{\underline{\underline{\bf #1}}}

\def\r#1{(\ref{#1})}
\def\l#1{\label{#1}}
\def\c#1{\cite{#1}}

\def\le{\left(}
\def\ri{\right)}
\def\les{\left[}
\def\ris{\right]}
\def\lec{\left\{}
\def\ric{\right\}}

\def\.{\mbox{ \tiny{$^\bullet$} }}

\def\eps{\epsilon}

\def\epsa{\epsilon_a}
\def\epsb{\epsilon_b}

\begin{document}
\bibliographystyle{unsrt}

\LARGE
\begin{center}
{\bf On the Bergman--Milton bounds for the homogenization of
dielectric composite materials}

\vspace{10mm} \large

Andrew J. Duncan\footnote{Corresponding Author. Fax: + 44 131 650
6553; e--mail: Andrew.Duncan@ed.ac.uk.},  Tom G.
Mackay\footnote{Fax: + 44 131
650 6553; e--mail: T.Mackay@ed.ac.uk.}\\
{\em School of Mathematics,
University of Edinburgh, Edinburgh EH9 3JZ, UK}\\
\vspace{3mm}
 Akhlesh  Lakhtakia\footnote{Fax:+1 814 865 99974; e--mail: akhlesh@psu.edu}\\
 {\em CATMAS~---~Computational \& Theoretical
Materials Sciences Group\\ Department of Engineering Science and
Mechanics\\ Pennsylvania State University, University Park, PA
16802--6812, USA}

\end{center}

\vspace{4mm}

\normalsize

\begin{abstract}
The Bergman--Milton bounds provide limits on the
effective  permittivity of a  composite material comprising
 two  isotropic dielectric materials. These provide tight bounds
for composites arising from many conventional materials. We
reconsider the Bergman--Milton bounds in light of the recent
emergence of metamaterials, in which unconventional parameter ranges
for relative permittivities are encountered. Specifically, it is
demonstrated that: (a) for nondissipative materials the bounds may
be unlimited if the constituent materials have relative
permittivities of opposite signs; (b) for weakly dissipative
materials characterized by relative permittivities with real parts
of opposite signs, the bounds  may be exceedingly large.

\end{abstract}

\noindent {\bf Keywords:} Bergman--Milton bounds; Maxwell Garnett
estimates; Hashin--Shtrikman bounds; metamaterials

\section{Introduction}

Increasingly, new materials which exhibit novel and potentially
useful electromagnetic responses are being developed \c{CM_03,CM_00}. At the
forefront of this rapidly expanding field lie \emph{metamaterials} \c{Walser}.
 These
are artificial composite materials which exhibit properties that
are either not exhibited
 by their constituents at all, or not exhibited to the same extent by
their constituents. With the emergence of these new
materials~---~which may exhibit radically different
properties to those encountered  traditionally in
electromagnetics/optics~---~some re--evaluation of established
theories is necessary. A prime example is provided by the recent
development of metamaterials which support planewave propagation
with negative phase velocity \c{FNR}. The experimental demonstration of
negative refraction in 2000 prompted  an explosion of interest in
issues  pertaining to negative phase velocity and negative
refraction \c{BKV,SAR}.

The process of homogenization, whereby two (or more) homogeneous
constituent materials are blended together to produce a composite
material which is effectively homogeneous within the
long--wavelength regime, is an important vehicle in the
conceptualization of metamaterials \c{Mackay05}. The estimation of
the effective constitutive parameters of homogenized composite
materials (HCMs) is a well--established process \c{L96}, aspects of
which have been revisited recently in light of the development of
exotic materials that exhibit properties such as negative phase
velocity. For example, it was demonstrated that two widely used
homogenized formalisms, namely the Maxwell Garnett and Bruggeman
formalisms, do not provide useful estimates of the HCM permittivity
within certain  parameter regimes \c{ML_04}. The Maxwell Garnett
estimates coincide with the well--known Hashin--Shtrikman bounds
\c{HS} on the HCM permittivity. While the former are commonly
implemented for both dissipative and nondissipative HCMs, the later
were derived for nondissipative HCMs.


In view of the limitations of  the Maxwell Garnett and Bruggeman
formalisms within certain parameter regimes, we explore in this
communication the implementation of the Bergman--Milton bounds
\c{Bergman_PRB,Milton_JAP,Bergman_PRL} for these parameter regimes.
 To be specific, we consider the
homogenization of two isotropic dielectric constituent materials
with relative permittivities $\eps_a$ and $\eps_b$. We explore the
regime in which the parameter\footnote{$\mbox{Re} \le \eps_{a,b}
\ri$ and $\mbox{Im} \le \eps_{a,b} \ri$ denote the real and
imaginary parts of $\eps_{a,b}$, respectively; $\mathbb{R}$ and
$\mathbb{C}$ denote the sets of real and complex numbers,
respectively. }
\begin{equation}
 \delta = \frac{\mbox{Re} \le \eps_a \ri }{\mbox{Re} \le \eps_b \ri}\,, \qquad
 \qquad   (\eps_a, \eps_b \in
 \mathbb{C})\,, \l{delta}
\end{equation}
is negative--valued, as this is where the Maxwell Garnett and
Bruggeman estimates are not useful \c{ML_04}. Notice that the
definition \r{delta} caters to
   the possibility that
 only one of
$\eps_{a}$ or $\eps_{b} \in \mathbb{R}$, as might arise for a
metal--in--insulator HCM, for example.
 In fact, the $\delta < 0$
regime which occurs for metal--in--insulator HCMs
\c{Aspnes,Milton_APL} is highly pertinent to the homogenization of
HCMs which support planewave propagation with negative phase
velocity.

Let us note that, although  the $\delta<0$ regime has been
discussed in the past in the context of the Bergman--Milton bounds
\c{Milton_JAP,Milton_APL,Milton81}, the discussion on the
inadequacy of those bounds has been brief. Amplification is needed
because of the possibility of fabricating negatively refracting
composite materials \c{ML05,ML06}, for example.

\section{Bergman--Milton bounds}

Two  bounds on the effective relative permittivity $\eps_{e}$ of the
chosen composite material  were established by Bergman
\c{Bergman_PRB,Bergman_Phys_Rep,Bergman_annals_phys,Bergman_SIAM}
and Milton \c{Milton_JAP,Miltonbook}. We write these as
$\mathsf{BM}_{\alpha}$ and $\mathsf{BM}_{\beta}$. In terms of a
real--valued parameter $\gamma$, these are given by (see  eqn. (24)
in \c{Milton_JAP})
\begin{equation}
\label{miltonbound1} \mathsf{BM}_{\alpha}(\gamma) = f_a \epsilon_a
+ f_b \epsilon_b - \frac{f_a f_b (\epsilon_b-\epsilon_a)^2}{3 \les
\gamma \epsilon_a + \le 1-\gamma \ri \epsilon_b \ris } \,, \qquad
 \qquad   (\eps_a, \eps_b \in
 \mathbb{C})\,,
\end{equation}
and
\begin{equation} \label{miltonbound2}
\mathsf{BM}_{\beta}(\gamma) = \lec \frac{f_a}{\epsilon_a} +
\frac{f_b}{\epsilon_b} - \frac{ 2f_a f_b \le \epsilon_a -
\epsilon_b \ri^2}{3  \, \eps_a \eps_b \les \epsilon_b  \gamma +
\eps_a \le 1 - \gamma  \ri \ris} \ric^{-1},\qquad
 \qquad   (\eps_a, \eps_b \in
 \mathbb{C})\,,
\end{equation}
where $f_{a,b}$   denotes the volume fraction of  the
constituent material with relative permittivity $\eps_{a,b}$,
and $f_a + f_b = 1$. For the bound $\mathsf{BM}_{\alpha}$
the parameter $\gamma$ takes the values $ \le 1 - f_a \ri /3 \leq
\gamma \leq 1 - f_a  /3$, whereas for the bound
$\mathsf{BM}_{ \beta}$ the parameter $\gamma$ takes the values $ 2
\le 1-  f_a \ri /3 \leq \gamma \leq 1 -  2f_a /3$.

The Bergman--Milton bounds \r{miltonbound1} and \r{miltonbound2} are
related to the two
 Maxwell Garnett estimates of the HCM relative permittivity
\c{L96,Aspnes}
\begin{equation} \label{MGa}
\mathsf{MG}_\alpha=\epsilon_b+\frac{3f_a
\epsilon_b(\epsilon_a-\epsilon_b)}
{\epsilon_a+2\epsilon_b-f_a(\epsilon_a-\epsilon_b)}, \qquad
 \qquad   (\eps_a, \eps_b \in
 \mathbb{C})\,,
\end{equation}
\begin{equation} \label{MGb}
\mathsf{MG}_\beta=\epsilon_a+\frac{3f_b
\epsilon_a(\epsilon_b-\epsilon_a)}
{\epsilon_b+2\epsilon_a-f_b(\epsilon_b-\epsilon_a)}, \qquad
 \qquad   (\eps_a, \eps_b \in
 \mathbb{C})\,.
\end{equation}
The Maxwell Garnett estimates represent the extension of the
Hashin--Shtrikman bounds \c{HS} into the complex--valued
permittivity regime. For nondissipative HCMs, the Maxwell Garnett
estimates coincide with the Bergman--Milton bounds  when the
parameter $\gamma$ attains
 its minimum and maximum   values; i.e.,
\begin{equation}
\left.
\begin{array}{lllll}
\displaystyle{ \mathsf{BM}_{ \alpha} \le \frac{1-f_a}{3} \ri } &=&
\displaystyle{\mathsf{BM}_{ \beta} \le \frac{2 -2 f_a}{3} \ri } &= &
\mathsf{MG}_{\alpha} \\ &&&& \\ \displaystyle{\mathsf{BM}_{ \alpha}
\le 1 - \frac{f_a}{3} \ri} & =& \displaystyle{ \mathsf{BM}_{ \beta}
\le 1 - \frac{2 f_a}{3} \ri} & = & \mathsf{MG}_{\beta}
\end{array}
\right\} \,. \l{HS-BM}
\end{equation}

In view of our particular interest in homogenization scenarios for
which $\delta < 0$, we note that
\begin{equation}
\left| \mathsf{BM}_{ \alpha} \le \frac{1- f_a}{3} \ri \right| =
\left| \mathsf{BM}_{ \beta} \le \frac{2 -2 f_a}{3} \ri  \right|=
\left| \mathsf{MG}_{ \alpha}  \right| \rightarrow \infty \qquad
\mbox{as} \qquad \delta \rightarrow \frac{f_b - 3}{f_b}\,
\l{HSa_lim}
\end{equation}
and
\begin{equation}
\left| \mathsf{BM}_{ \alpha} \le 1 -  \frac{ f_a}{3} \ri \right|
  = \left|  \mathsf{BM}_{ \beta} \le 1 - \frac{2 f_a}{3} \ri
\right| = \left| \mathsf{MG}_{ \beta}  \right| \rightarrow \infty
\qquad \mbox{as} \qquad \delta \rightarrow \frac{f_a}{f_a - 3} \,
\l{HSb_lim}
\end{equation}
for nondissipative mediums.
Thus, there exist
\begin{itemize}
\item[(i)] a volume fraction $f_a \in \le 0,1 \ri$ at which
$\mathsf{MG}_\alpha$ is unbounded for
  all values of $\delta < -2$, and
  \item[(ii)]  a volume fraction $f_a \in \le 0,1 \ri$ at which
$\mathsf{MG}_\beta$ is unbounded for
  all values of $\delta \in \le -1/2,  0 \ri $.
  \end{itemize}

\section{Numerical illustrations}

Let us now numerically explore    the Bergman--Milton bounds, along
with the Maxwell Garnett estimates, for some illustrative examples
of nondissipative and dissipative HCMs. The parameter $\delta$,
defined in \r{delta},  is used to classify the two constituent
materials of the chosen HCMs. We begin in \S3.1 by considering
nondissipative HCMs. While these do not represent realistic
materials, they provide valuable insights into the limiting process
in which weakly dissipative materials become nondissipative.
Furthermore, they provide a useful yardstick in the evaluation of
dissipative HCMs, which are considered in \S3.2.

\subsection{Nondissipative HCMs}

We begin with the most straightforward situation:
nondissipative HCMs arising from constituent materials with
$\delta
> 0$. In Figure~1,
 the Maxwell Garnett estimates
$\mathsf{MG}_{\alpha}$ and $\mathsf{MG}_{\beta}$ (which in this case
are identical to the  Hashin--Shtrikman bounds) are plotted against
$f_a \in (0,1)$ for $\epsa=6$ and $\epsb=2$. The Bergman--Milton
bound $\mathsf{BM}_{\alpha}$ is given for $f_a \in \lec 0.1, 0.2,
0.3, 0.4, 0.5, 0.6, 0.7, 0.8, 0.9 \ric $. The corresponding plots
 of $\mathsf{BM}_{\beta}(\gamma)$ with $\gamma$ overlies that of
$\mathsf{BM}_{\alpha}$.  The Bergman--Milton bounds are entirely
contained within the envelope constructed by the Maxwell Garnett
estimates.

Let us turn now to the nondissipative scenario wherein $\delta < 0$.
In Figure~2, the
 the Maxwell Garnett estimates
$\mathsf{MG}_{\alpha}$ and $\mathsf{MG}_{\beta}$ are presented as
functions of $f_a$ for $\epsilon_a=-6$ and $\epsilon_b=2$.
 The Bergman--Milton bound $\mathsf{BM}_{ \alpha}$ is given for $f_a \in
\lec 0.1, 0.2, 0.3, 0.4, 0.5, 0.6, 0.7, 0.8, 0.9 \ric $. The
corresponding Bergman--Milton bound $\mathsf{BM}_{ \beta}$ is
plotted in Figure~3. In consonance with    \r{HS-BM} and
\r{HSa_lim}, we see that $\mathsf{MG}_\alpha$ becomes unbounded as
$f_a \rightarrow 0.25$. It is clear  that   $\mathsf{MG}_{\beta}
\leq \mathsf{BM}_{\alpha} \leq \mathsf{MG}_{\alpha}$ for $f_a <
0.25$, whereas $\mathsf{MG}_{\alpha} \leq \mathsf{BM}_{\beta} \leq
\mathsf{MG}_{\beta}$ for $f_a > 0.25$. For $f_a > 0.25$, the
Bergman--Milton bound $\mathsf{BM}_{\alpha}$ lies outside both
 Maxwell Garnett estimates
$\mathsf{MG}_{\alpha}$ and $\mathsf{MG}_{\beta}$, and similarly
$\mathsf{BM}_{\beta}$ lies outside both
 Maxwell Garnett estimates
$\mathsf{MG}_{\alpha}$ and $\mathsf{MG}_{\beta}$
 for $f_a < 0.25$, although the relations
\r{HS-BM}  still hold.

\subsection{Dissipative HCMs}

We turn to homogenization scenarios  based on  dissipative
constituent materials; i.e., $\epsilon_{a,b}\in\mathbb{C}$. Let us
begin with the $\delta > 0$ scenario. In Figure~4, the
homogenization of constituents characterized by the relative
permittivities  $\epsilon_a=6+0.3i$ and $\epsilon_b=2+0.2i$ is
illustrated. In this figure, the Maxwell Garnett estimates on
complex--valued $\epsilon_{e}$ are plotted  as $f_a$ varies from
$0$ to $1$. The Bergman--Milton bounds, which  are graphed for
$f_a \in \lec 0.1, 0.2, 0.3, 0.4, 0.5, 0.6, 0.7, 0.8, 0.9 \ric $,
are fully contained within the Maxwell Garnett envelope. That is,
we have $\mathsf{MG}_{\beta} \leq \mathsf{BM}_{\alpha, \beta} \leq
\mathsf{MG}_{\alpha}$ for all values of $f_a$.

 Now we  consider dissipative constituent materials with
 $\delta < 0$. In Figure~5, the homogenization of
 constituent materials given by $\epsilon_a=-6+3i$ and
$\epsilon_b=2+2i$ is represented. The Maxwell Garnett estimates
are plotted for $f_a \in (0,1)$,  whereas the
 Bergman--Milton bounds are given for
$f_a \in \lec 0.1, 0.2, 0.3, 0.4, 0.5, 0.6, 0.7, 0.8, 0.9 \ric $.
As is the case in Figure~4, $\mathsf{BM}_\beta$ lies entirely
within the envelope constructed by $\mathsf{MG}_\alpha$ and
$\mathsf{MG}_\beta$. We see that $\mathsf{BM}_\alpha \ge
\mathsf{MG}_{\beta}$ for all values of $f_a$; but, for mid--range
values of $f_a$, $\mathsf{BM}_\alpha$
 slightly exceeds
$\mathsf{MG}_\alpha$ for certain values of the parameter $\gamma$.

As  the degree of dissipation exhibited by the constituent
materials is decreased, the extent to which  $\mathsf{BM}_\alpha$
exceeds $\mathsf{MG}_\alpha$ is increased. This is illustrated in
Figure~6  wherein the homogenization  is repeated  with
$\epsilon_a=-6+i$ and $\epsilon_b=2+2i/3$. As in Figure~4, the
Maxwell Garnett estimates are plotted for $f_a \in (0,1)$, while
the
 Bergman--Milton bounds are given for
$f_a \in \lec 0.1, 0.2, 0.3, 0.4, 0.5, 0.6, 0.7, 0.8, 0.9 \ric $.
The Bergman--Milton bound $\mathsf{BM}_\beta$ lies within the
Maxwell Garnett  envelope for all values of $f_a$, but substantial
parts of $\mathsf{BM}_\alpha$ lie well outside the envelope of the
two Maxwell Garnett estimates.

The behaviour observed in Figures~5 and 6 is further exaggerated
in Figure~7, where the homgenization of constituent materials with
$\epsilon_a=-6+0.3i$ and $\epsilon_b=2+0.2i$ is represented. The
Maxwell Garnett estimates are plotted for $f_a \in (0,1)$; for
reasons of clarity, the
 Bergman--Milton bounds are plotted only for
$f_a \in \lec 0.1,  0.3,  0.5 \ric $. The Maxwell Garnett
estimates are exceedingly large and the Bergman--Milton bounds
are larger still.

Finally, let us focus on the scenario referred to in the
introduction, namely the homogenization of a  conducting
constituent material
and a nonconducting constituent material, wherein $\delta < 0$.
Suppose we consider  constituents characterized by the  relative
permittivities $\epsilon_a=-6+3i$ and $\epsilon_b=2$. In Figure~8
the Maxwell Garnett estimates are plotted for $f_a \in (0,1)$,
whereas the
 Bergman--Milton bounds are given for
$f_a \in \lec 0.1, 0.2, 0.3, 0.4, 0.5, 0.6, 0.7, 0.8, 0.9 \ric $.
As we observed in Figure~6,
 the Maxwell Garnett envelope does not contain substantial parts
 of the
Bergman--Milton  bound $\mathsf{BM}_\alpha$, whereas the
$\mathsf{BM}_\beta$ bound lies entirely within the envelope
constructed from  the two Maxwell Garnet estimates.

\section{Discussion and conclusions}

The Bergman--Milton bounds, as well as the Maxwell Garnett
estimates,  are valuable for estimating the effective constitutive
parameters of HCMs in many commonly encountered circumstances.
However, the advent of exotic new materials and metamaterials has
led to the examination of such bounds within unconventional
parameter regimes. It was recently demonstrated that  the Bruggeman
homogenization formalism and the Maxwell Garnett homogenization
formalism  do not provide useful estimates of the HCM  permittivity
when the relative permittivities of the constituent materials
$\epsilon_a$ and $\epsilon_b$ are such that \c{ML_04}
\begin{itemize}
 \item[(i)]
 $\mbox{Re} \le \eps_a \ri$ and $\mbox{Re} \le
\eps_b \ri$ have opposite signs;  and \item[(ii)] $| \mbox{Re} \le
\eps_{a,b} \ri |$ $ \gg $ $| \mbox{Im} \le \eps_{a,b} \ri |$.
\end{itemize}
Similarly, we have demonstrated in the preceding sections of this
communication, that the Bergman--Milton bounds do not provide tight
limits on the value of $\epsilon_{e}$ within the same parameter
regime.

We note that if the real parts of $\eps_a$ and $\eps_b$ have
opposite signs, but  are of the same order of magnitude as their
imaginary parts, then the Bergman--Milton bounds are indeed useful,
and they then lie within the envelope constructed by the
Maxwell Garnett estimates.\\

\vspace{10mm}

\noindent {\bf Acknowledgement:}
AL is grateful for many discussions
with Bernhard Michel of Scientific Consulting, Rednitzhembach, Germany.
The authors thank anonymous referees for their helpful remarks.
\\

\newpage

\begin{figure}
\begin{center}
\resizebox{5in}{!}{\includegraphics{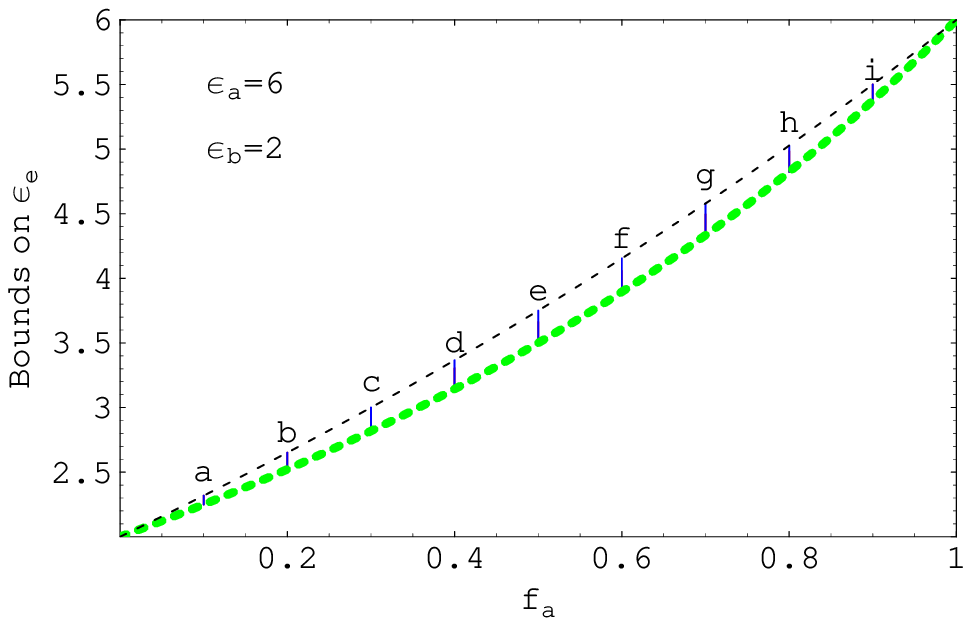}}
\caption{\footnotesize{The
 $\mathsf{MG}_{\alpha}$ (thick dashed line) and  $\mathsf{MG}_{\beta}$ (thin
    dashed line)
 estimates of $\epsilon_{e}$ plotted
against
    $f_a$  for $\epsilon_a=6$, $\epsilon_b=2$.
  The vertical solid lines represent the variation
of the Bergman--Milton bound $\mathsf{BM}_{\alpha}$ with $\gamma$
for
$f_a\in\{0.1(a),0.2(b),0.3(c),0.4(d),0.5(e),0.6(f),0.7(g),0.8(h),0.9(i)\}$;
and these  coincide with the corresponding variation of
$\mathsf{BM}_{\beta}$ with $\gamma$. }}
    \label{real_1}
\end{center}
\end{figure}

\newpage

\begin{figure}
\begin{center}
\resizebox{5in}{!}{\includegraphics{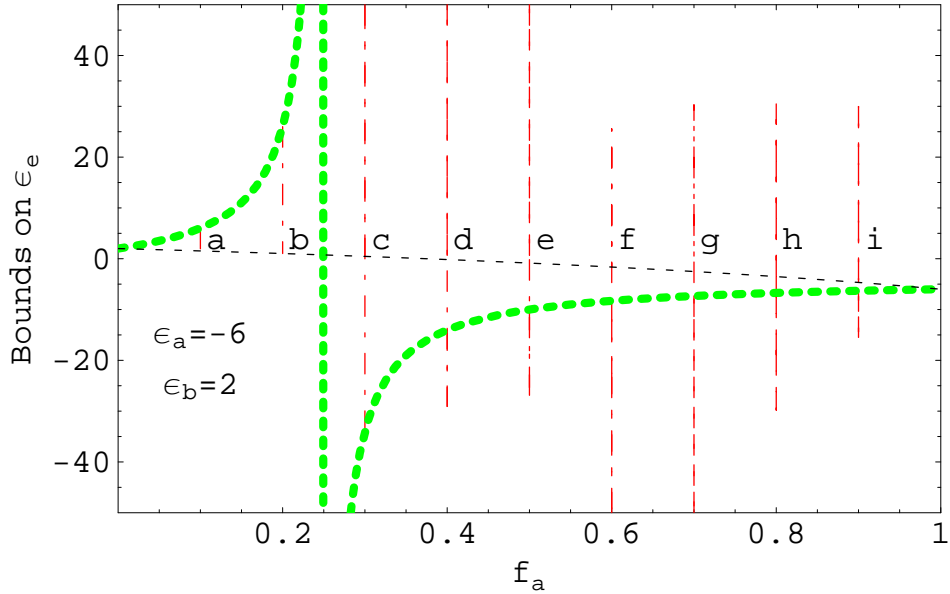}}
\caption{\footnotesize{The $\mathsf{MG}_{\alpha}$ (thick dashed
line) and  $\mathsf{MG}_{\beta}$ (thin dashed line)  estimates of
$\epsilon_e$ plotted against $f_a$ for $\epsilon_a=-6$ and
$\epsilon_b=2$.
 The Bergman--Milton
bound
$\mathsf{BM}_{\alpha}$ is plotted as the vertical broken lines for
$f_a\in\{0.1(a),0.2(b),0.3(c),0.4(d),0.5(e),0.6(f),0.7(g),0.8(h),0.9(i)\}$.}}\label{real_2a}
\end{center}
\end{figure}

\vspace{10mm}

\newpage
\begin{figure}
\begin{center}
\resizebox{5in}{!}{\includegraphics{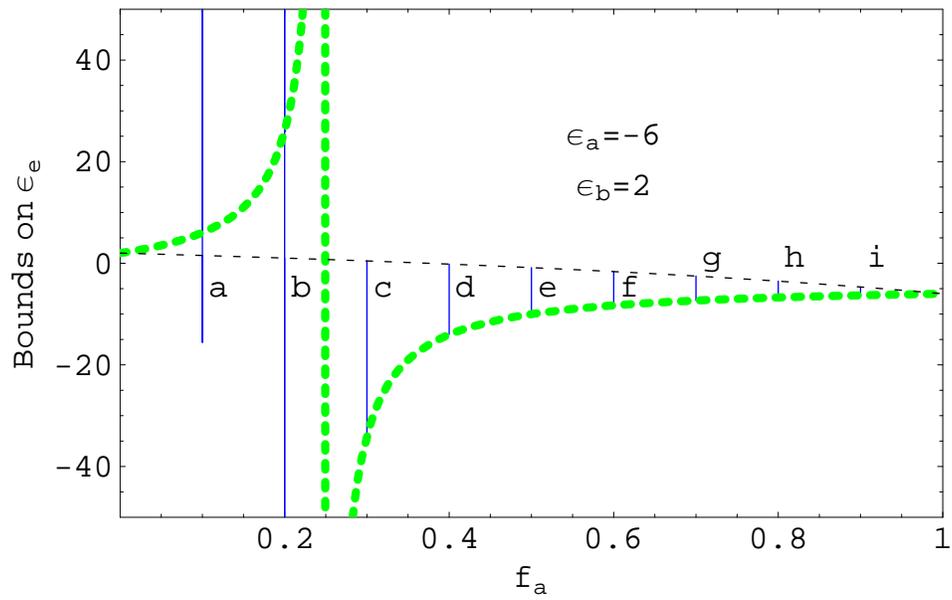}}
\caption{\footnotesize{As Figure~2 but with $\mathsf{BM}_{\beta}$
(vertical solid lines) in place of $\mathsf{BM}_{\alpha}$.}}
\label{real_2b}
\end{center}
\end{figure}

\newpage

\begin{figure}
\begin{center}
\resizebox{5in}{!}{\includegraphics{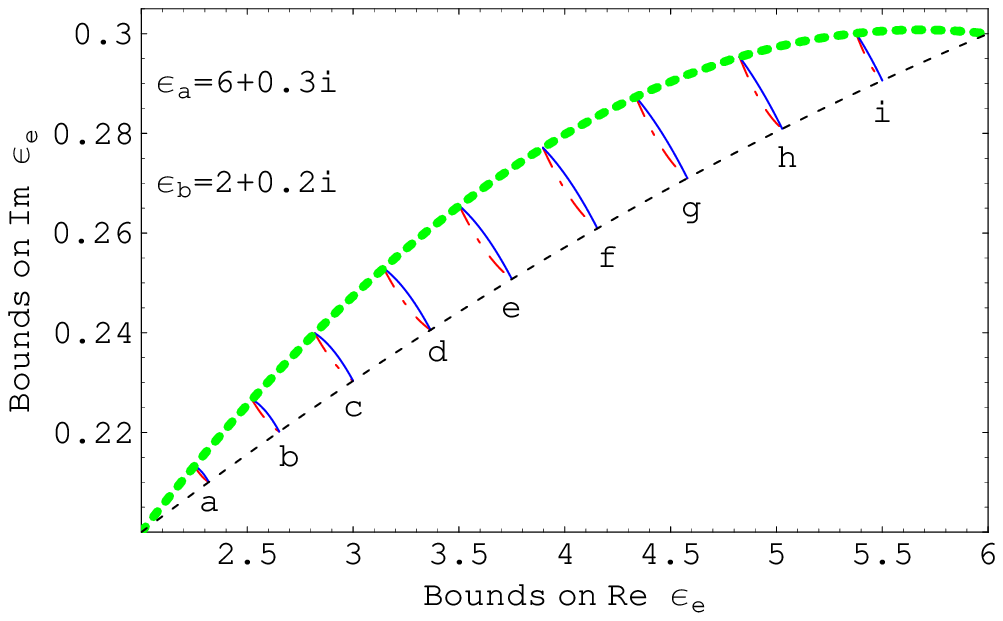}} \vspace{20mm}
\resizebox{5in}{!}{\includegraphics{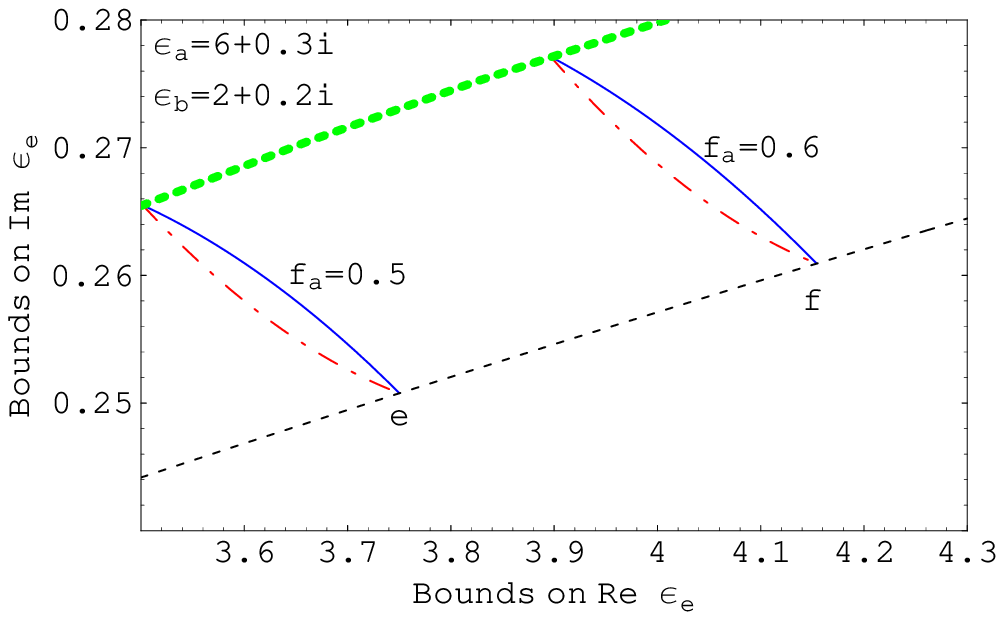}}
\caption{\footnotesize{ The $\mathsf{MG}_{\alpha}$ (thick dashed
line)
    and  $\mathsf{MG}_{\beta}$ (thin
    dashed line) estimates
in relation to  $\mbox{Re}\,\epsilon_{e}$ and
$\mbox{Im}\,\epsilon_{e}$
    as $f_a$ varies from $0$ to $1$,
    for $\epsilon_a=6+0.3i$,
    $\epsilon_b=2+0.2i$. The Bergman--Milton bounds
 $\mathsf{BM}_{\alpha}$ (thin broken dashed lines) and  $\mathsf{BM}_{\beta}$ (thin solid lines)
in the top diagram are plotted for
    $f_a\in\{0.1(a),0.2(b),0.3(c),0.4(d),0.5(e),0.6(f),0.7(g),0.8(h),0.9(i)\}$.
    The bottom diagram  shows the Bergman--Milton
    bounds in greater detail but  for $f_a=0.5(e)$ and $f_a=0.6(f)$.}}
\label{weak_1}
\end{center}
\end{figure}

\newpage

\begin{figure}
\begin{center}
\resizebox{5in}{!}{\includegraphics{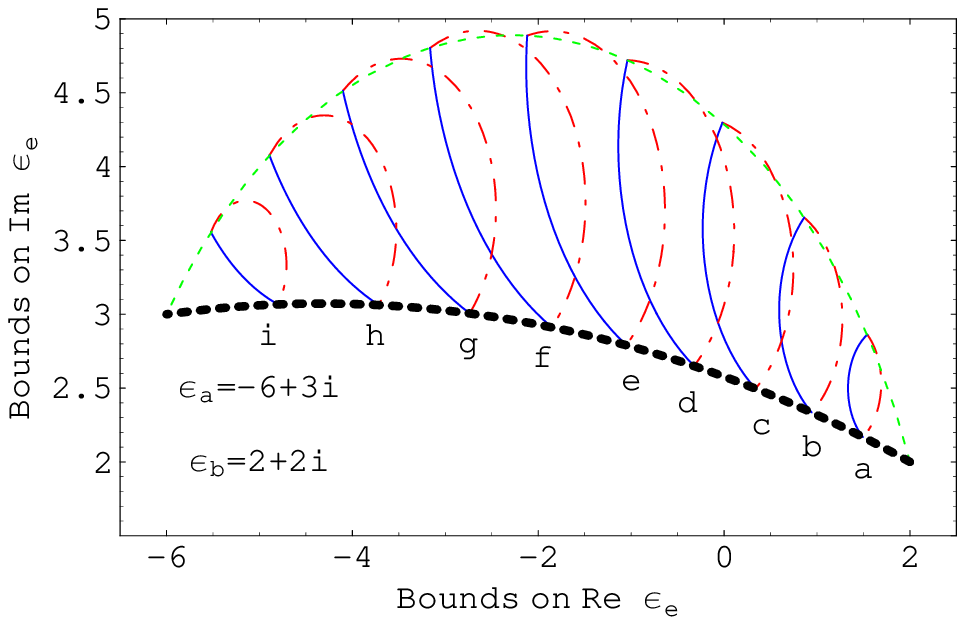}}
\caption{\footnotesize{ The $\mathsf{MG}_{\alpha}$ (thin dashed
line) and
    $\mathsf{MG}_{\beta}$ (thick
    dashed line) estimates
in relation to $\mbox{Re}\,\epsilon_{e}$ and $\mbox{Im}
\,\epsilon_{e}$
    as $f_a$ varies from $0$ to $1$,  for $\epsilon_a=-6+3i$ and
    $\epsilon_b=2+2i$. The Bergman--Milton bounds  $\mathsf{BM}_{\alpha}$
 (thin broken dashed lines) and $\mathsf{BM}_{\beta}$ (thin solid lines)
 are plotted for
    $f_a\in\{0.1(a),0.2(b),0.3(c),0.4(d),0.5(e),0.6(f),0.7(g),0.8(h),0.9(i)\}$.}}
\label{high_2}
\end{center}
\end{figure}

\newpage

\begin{figure}
\begin{center}
\resizebox{5in}{!}{\includegraphics{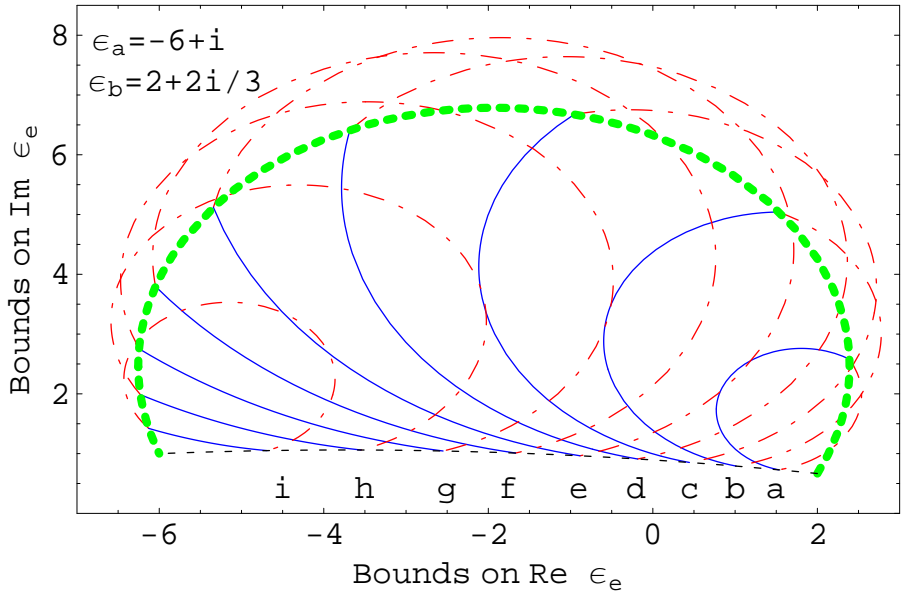}}
\caption{\footnotesize{As Figure \ref{weak_1} but for
    $\epsilon_a=-6+i$, $\epsilon_b=2+ 2i/3$.}}
\label{high_1}
\end{center}
\end{figure}

\newpage

\begin{figure}
\begin{center}
\resizebox{5in}{!}{\includegraphics{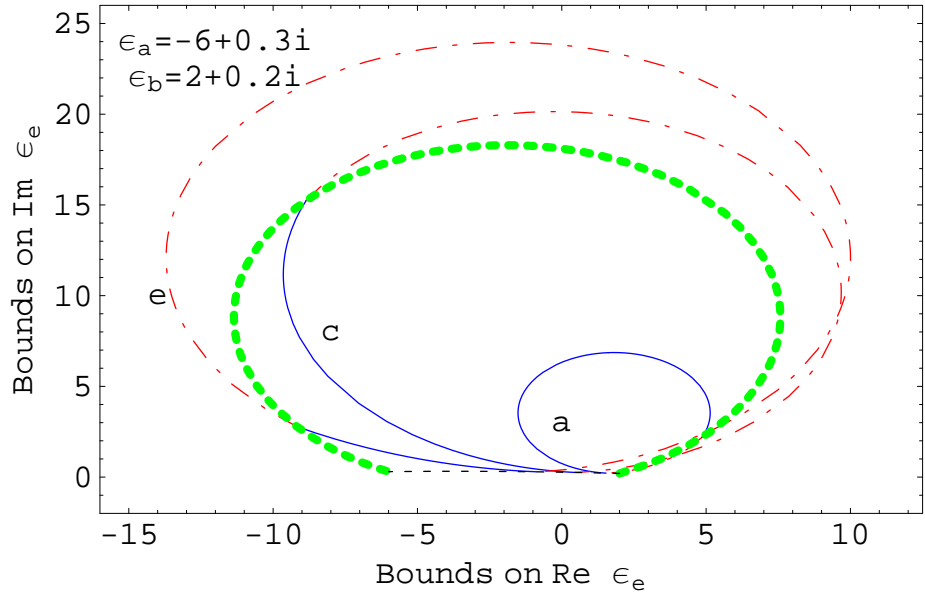}}
\caption{\footnotesize{As Fig. \ref{weak_1} but for
$\epsilon_a=-6+0.3i$ and $\epsilon_b=2+0.2i$. The Bergman--Milton
    bounds are  plotted for $f_a\in\{0.1(a),0.3(c),0.5(e)\}$.}}
\label{weak_2}
\end{center}
\end{figure}

\newpage

\begin{figure}
\begin{center}
\resizebox{5in}{!}{\includegraphics{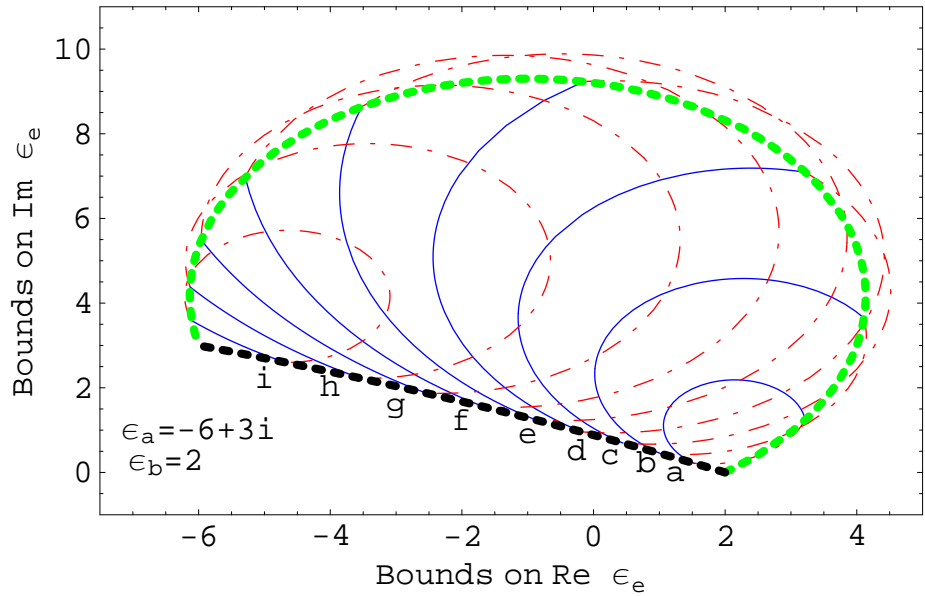}}
\caption{\footnotesize{As Fig. \ref{weak_1} but for
$\epsilon_a=-6+3i$ and $\epsilon_b=2$. }}
\label{metal_insulator}
\end{center}
\end{figure}

\end{document}